\documentclass[%
 reprint,
 amsmath,amssymb,
 aps,
 pra,
]{revtex4-1}

\usepackage{graphicx}
\usepackage{dcolumn}
\usepackage{bm}
\usepackage{nicefrac}

\usepackage{color}

\begin{document}

\preprint{APS/123-QED}

\title{Interference effect and Autler-Townes splitting in coherent blue light generated by four-wave mixing}

\author{M. P. Moreno}
\affiliation{Departamento de F\'{i}sica, Universidade Federal de Rond\^{o}nia, 76900-726, Ji-Paran\'{a}, Rond\^{o}nia, Brazil}
\author{A. A. C. de Almeida}

\author{S. S. Vianna}%
 \email{vianna@ufpe.br}
\affiliation{Departamento de F\'{i}sica, Universidade Federal de Pernambuco, 50670-901, Recife, Pernambuco, Brazil}

\date{\today}

\begin{abstract}
An unexpected interference effect between four-wave mixing excitation routes in rubidium vapor is observed using a combination of a cw diode laser and a 1 GHz femtosecond pulse train. The generated coherent blue light is analyzed by scanning the diode frequency and the repetition rate of the frequency comb. In both cases, the cw laser induces an Autler-Townes splitting and a large separation is observed in the doublet structure due to the configuration of copropagating fields. Remarkably, by scanning the repetition rate, a narrow peak appears as a signature of the interfering pathways. A non-perturbative treatment of the Bloch equations achieves good agreement with the experimental data, and demonstrates that the interfering effect appears only when both lasers are resonant with one- and two-photon transitions for the same atomic velocity group.
\begin{description}
\item[PACS numbers]
 \verb+32.80.Qk, 42.50.Hz, 78.47.nj+ 
\end{description}
\end{abstract}

\maketitle

\section{Introduction}
Four-wave mixing (FWM) is a nonlinear process associated with many applications. In hot or room-temperature atomic vapors, it has been used to generate quantum correlated beams \cite{Ma, Kim}, to store quantum memory and transfer orbital angular momentum between light beams \cite{Chopinaud, Offer}, to reduce the paraxial diffraction of light \cite{Katzir}, to generate slow light \cite{Arsenovic} and to obtain a single-photon source \cite{Ripka}, for instance.

Most of these experiments are performed with cw lasers, exploring the high power and the tunability near atomic resonances. Femtosecond lasers have also been employed, making use of the coherent temporal control technique \cite{Mukamel}. 
In recent years, advances in ultrafast lasers have enhanced the direct application of mode-locked femtosecond (fs) lasers, leading for instance to the development of a rapid multidimensional coherent spectroscopy with high spectral resolution \cite{Cundiff2017}. In this sense, the combination of a cw laser and a femtosecond pulse train allow us not only to probe the action of each laser but also to explore their different characteristics in nonlinear processes.

In this work, we extend our previous study of the coherent blue light generated in rubidium vapor, now using a 1 GHz fs pulse train along with a cw diode laser instead of a 100 MHz pulse train \cite{Lopez}. Under this high repetition rate  of the fs laser, the necessary condition for the coherent accumulation of population is very well fulfilled. In this context, a good description of the FWM process in the frequency domain can be achieved. Moreover, with 1 GHz frequency separation of the optical modes we can easily distinguish the blue signal generated by each mode of the frequency comb. As a consequence, the results appear to be very similar to those obtained using two cw diode lasers \cite{Akulshin2009}, as will be seen from the analysis of the excitation spectra. In particular, we observed the Autler-Townes splitting \cite{Autler1955} in the FWM signal by scanning the frequency of the strong beam in a configuration of  copropagating fields.

\begin{figure}[htbp]
\centering
\includegraphics[width=1\linewidth]{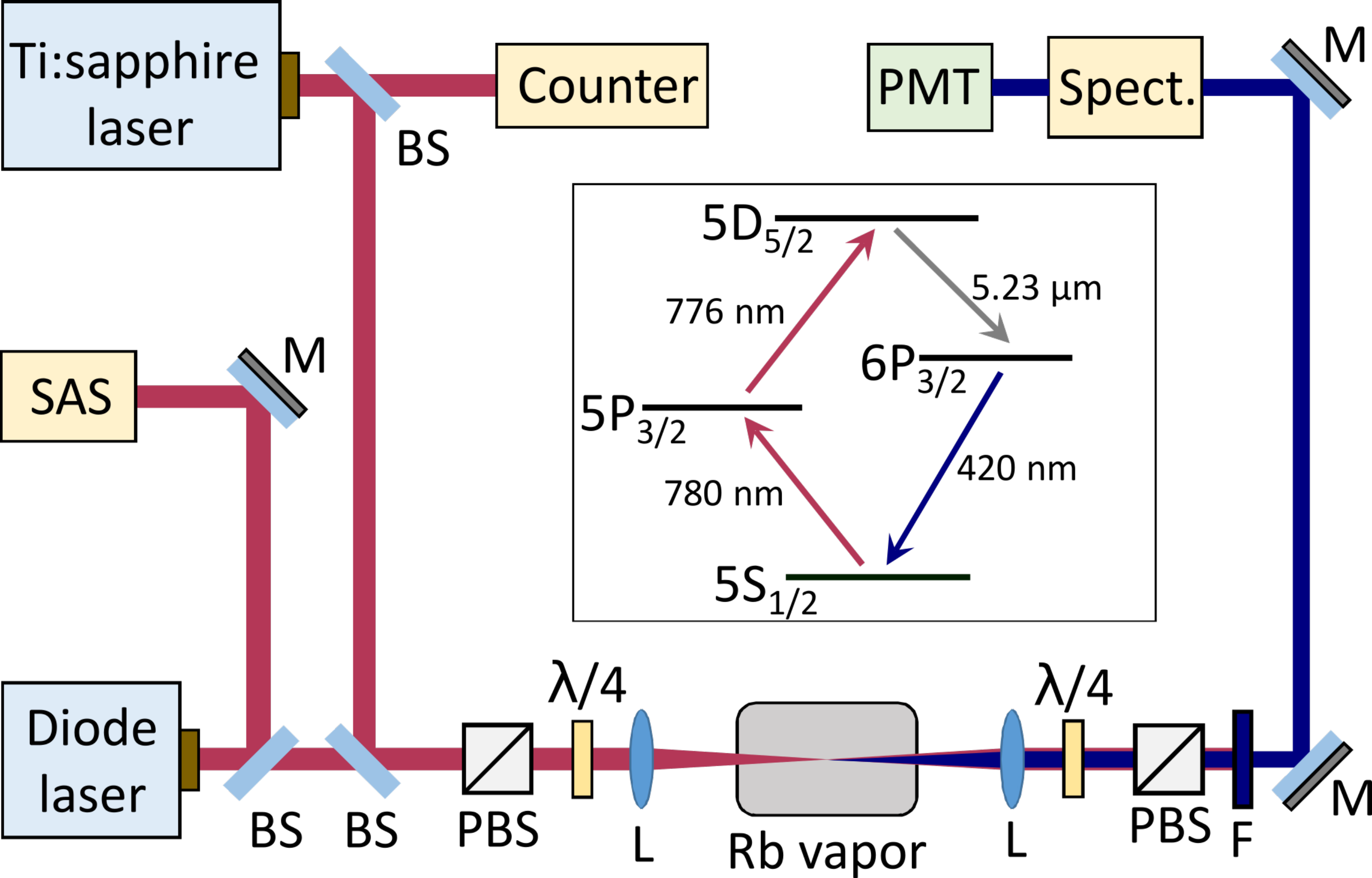}
\caption{(Color online) Experimental setup with relevant energy levels of $^{85}$Rb. BS, L, M, PBS, PMT and SAS indicate Beam-Splitter, Lens, Mirror, Polarizer Beam-Splitter, Photomultiplier and Saturation Absorption Spectroscopy, respectively.}
\label{fig1}
\end{figure}

The more interesting aspect of this experimental system is that the two-photon transition 5S$_{1/2} \rightarrow 5P_{3/2} \rightarrow 5$D can be driven by two routes: (i) two modes of the frequency comb and/or (ii) by the cw laser, in the 5S$_{1/2} \rightarrow 5P_{3/2}$ (780 nm) transition, while one of the frequency comb modes excites the 5P$_{3/2} \rightarrow 5$D (776 nm) transition. In the parametric FWM process investigated here, the nonlinear signal is determined by two-photon coherence between the 5S and 5D states, and by the amplified spontaneous emission at 5.2 $\mu$m~ \cite{Willis}. In this case, the generated blue light reflects not only the characteristics of the atomic system, but also carries the phase information related to the excitation beams. When the two excitation pathways occur for the same atomic velocity group, an interference effect can be clearly observed. The signature of this effect is a narrow peak over the Autler-Townes doublet. 

\section{Experimental setup and results}
Our experimental setup is schematically illustrated in Fig.~1. The fs pulse train is generated by a mode-locked Ti:sapphire laser (BR Labs Ltda), that emits 100 fs pulses at 1 GHz repetition rate, which is phase locked with 1-Hz resolution. The cw light source is a diode laser stabilized in temperature with a linewidth of about 1 MHz. The two beams, with parallel circular polarizations \cite{Akulshin2009}, copropagate through a 5 cm long cell containing natural Rb and heated up to $\approx 100\;^{o}$C.  The coherent blue light (CBL) is collected in the forward direction.

In the measurements, the average power of the fs laser was fixed at 250 mW at the cell entrance, while the power of the diode laser was varied between 0.07 and 4.5 mW. Both beams were focused at the center of the cell and then collimated. The 420 nm signal is selected using a blue bandpass filter and diffraction gratings and then sent to a photomultiplier tube and recorded by an oscilloscope.

\begin{figure}[htbp]
\centering
\includegraphics[width=0.9\linewidth]{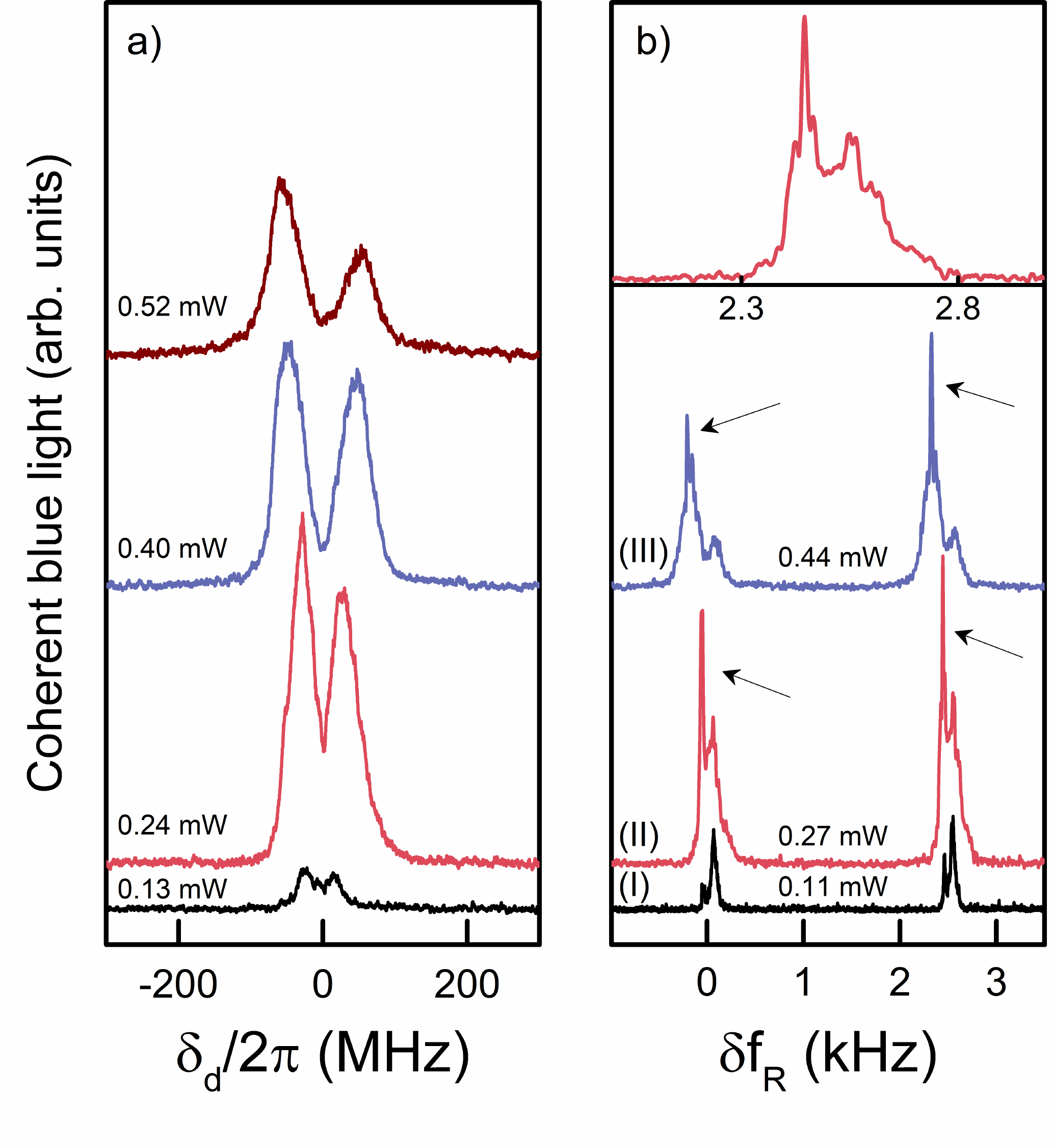}
\caption{(Color online) Coherent blue light intensity as a function of (a) diode frequency detuning (${\delta_{d}}/{2\pi}$), and (b) repetition rate variation ($\delta f_{R}$), for different powers of the diode laser as indicated in each curve. In (a) $f_{R}$ is fixed at $\sim$ 987.791 MHz while in (b) the diode frequency is fixed near the closed transition and $\delta f_{R}$ = 0 for $f_{R}$ = 987.749 886 MHz. The upper curve in (b) is a zoom of the right structure in curve (II).}
\label{fig2}
\end{figure}

Our results were focused at the isotope $^{85}$Rb, where we analyzed the CBL behavior with respect to the detuning and power of the diode laser, and repetition rate ($f_{R}$) of the pulse train. In the Fig. 2(a) we have the CBL as a function of the diode frequency detuning with respect to the 5S$_{1/2}, F=3 \rightarrow$ 5P$_{3/2}, F=4$ closed transition, with one mode of the frequency comb near to the 5P$_{3/2}, F=4 \rightarrow$ 5D$_{5/2}, F=5$ transition, for a locked $f_R \sim 987.791$ MHz. The cell temperature was set at $T \approx80$ $^o$C, corresponding to an atomic density of order of $10^{12}$ atoms/cm$^{3}$ \cite{Alcock}, and different powers of the diode laser were considered. We easily notice a doublet structure where the separation between the peaks depends on the power of the diode laser. This Autler-Townes (AT) splitting, observed when we scan the strong field frequency, is a kind of self-dressing effect \cite{Xiao2010}.

\begin{figure}[htbp]
\centering
\includegraphics[width=0.8\linewidth]{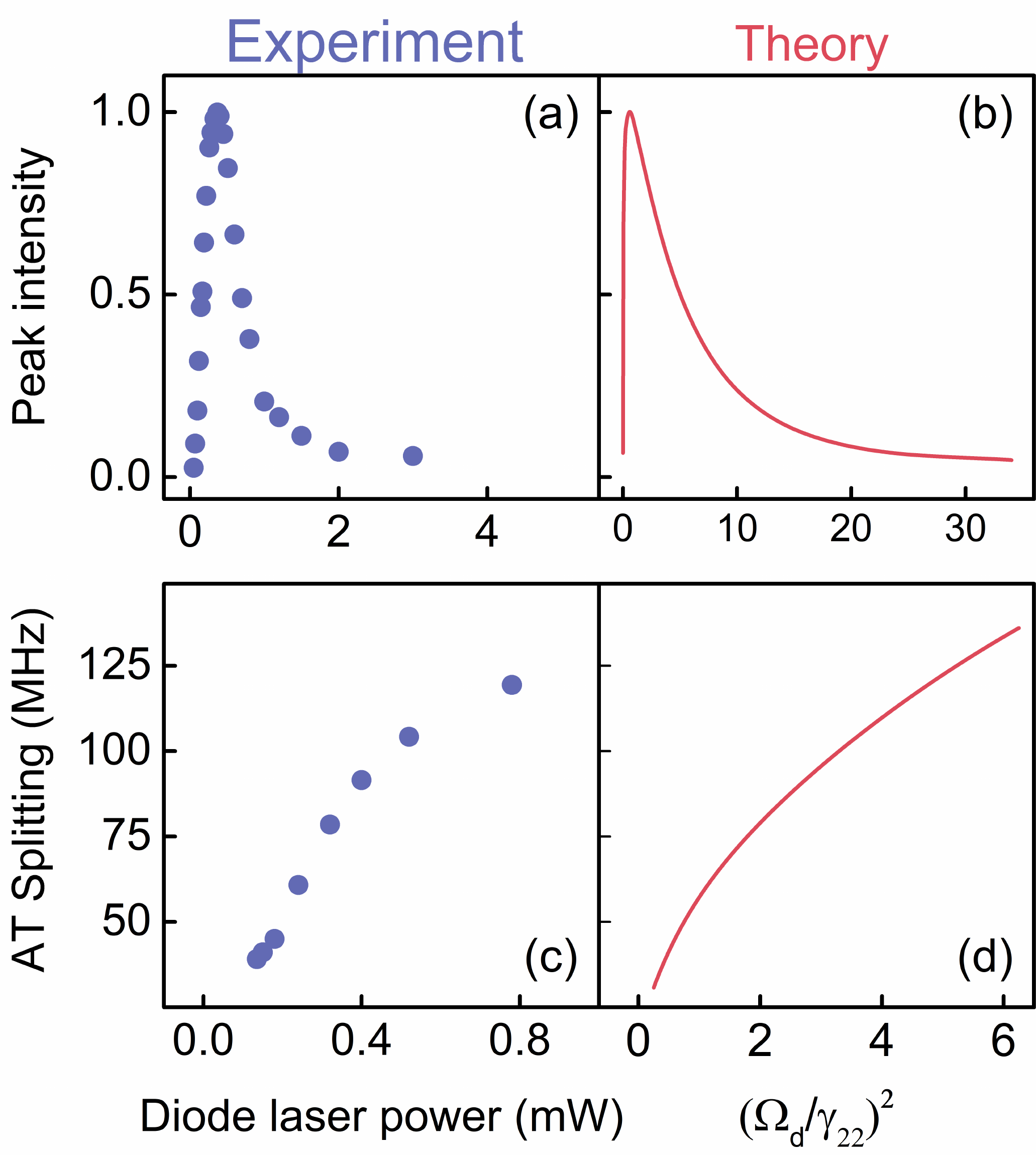}
\caption{(Color online) Comparison between experimental data ((a) and (c)) and numerical calculation ((b) and (d)) for the CBL peak intensity ((a) and (b)) and the Autler-Townes (AT) splitting ((c) and (d)) as a function of the diode laser power. $\gamma_{22}$ is the decay rate of the 5P$_{3/2}$ state.}
\label{fig3}
\end{figure}

The intensity dependence of the coherent blue signal with the diode power (measure at the entrance of the cell) is represented by the blue dots of Fig. 3(a). The experimental results were obtained for a fixed value of $f_R$, with a frequency mode near the closed transition 5P$_{3/2}, F=4 \rightarrow$ 5D$_{5/2}, F=5$, and for a cell temperature of $T = 85\;^o$C. Each point corresponds to an average of five scans and represents the higher peak value of one doublet. It is interesting to note that, for this experimental condition, the signal is maximum at $\approx 0.6$ mW and goes to zero for powers greater than 2 mW. The decrease of the signal for high powers of the diode is due to the ac Stark shift: as the diode power increases, the ac
Stark shift and therefore the detuning of the two-photon
resonance also increases, fading the signal. Another distinct feature are the large values observed for the doublet separation as a function of the diode power, displayed in Fig. 3(c) for the same experimental conditions of Fig. 2(a). These values are in contrast with a separation of the order of the Rabi frequency, usually found when the frequency of the strong field is fixed near resonance and the AT splitting is probed by the weak beam \cite{Verkerk1986}. This apparent contradiction will be discussed in Section III.

The dependence of the coherent blue light on the repetition rate is shown in Fig. 2(b). In this case, the diode frequency was tuned near to the 5S$_{1/2}, F=3 \rightarrow$ 5P$_{3/2}, F=4$ closed transition, while $f_R$ was scanning around $f_R^0 = 987.749\;886$ MHz, with $T = 74\;^o$C.  We represent the results for three diode powers, where the peaks, induced by the diode laser in combination with two neighboring modes of the frequency comb, are separated by $\approx 2.5$ kHz \cite{Moreno2011a}, which corresponds to $\approx 987$ MHz in the frequency of these modes. The doublet structure is also present with similar characteristics to those observed when the diode frequency is scanned. However, an intriguing feature is a narrow peak that appears superimposed on the broader AT peaks as indicated by the arrows in Fig. 2(b). This narrow peak is observed depending on the value of the repetition rate and a close-up of the doublet structure at the right of curve (II) is displayed on the upper curve in Fig. 2(b).

\section{Theory and discussion}
In order to explain the main features observed in the experiment, the Autler-Townes splitting  and the very narrow peak that appears when the repetition rate is scanning, we use a simple model consisting of independent four-level diamond-type systems interacting with three cw fields, as schematized in Fig. 4. Just as it is in the experiment, the first two transitions can be driven by different routes: (i) two modes of the frequency comb, $\omega_{n}$ and $\omega_{m}$ (\textit{fs-fs} pathway) and/or (ii) by the cw laser and one of the frequency comb modes, $\omega_{d}$ and $\omega_{m}$ (\textit{cw-fs} pathway). We have also included a seed for the 5D$_{5/2}$ $\rightarrow$ 6P$_{3/2}$ transition, a necessary condition to start the four wave mixing process and then to generate the blue beam.

\begin{figure}[htbp]
\centering
\includegraphics[width=0.7\linewidth]{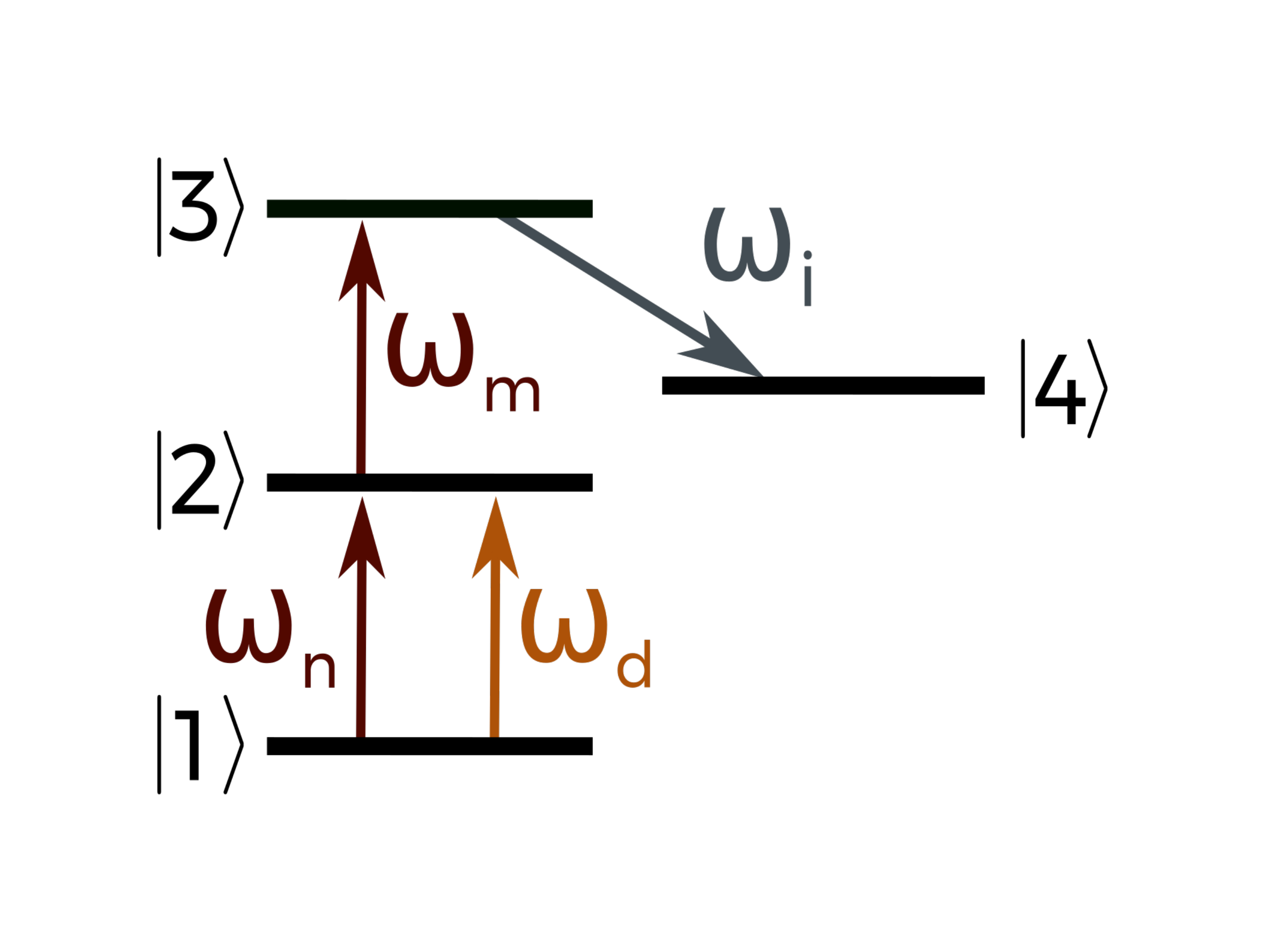}
\caption{(Color online) Four-level theoretical model.}
\label{fig4}
\end{figure}

\begin{figure}[htbp]
\centering
\includegraphics[width=1\linewidth]{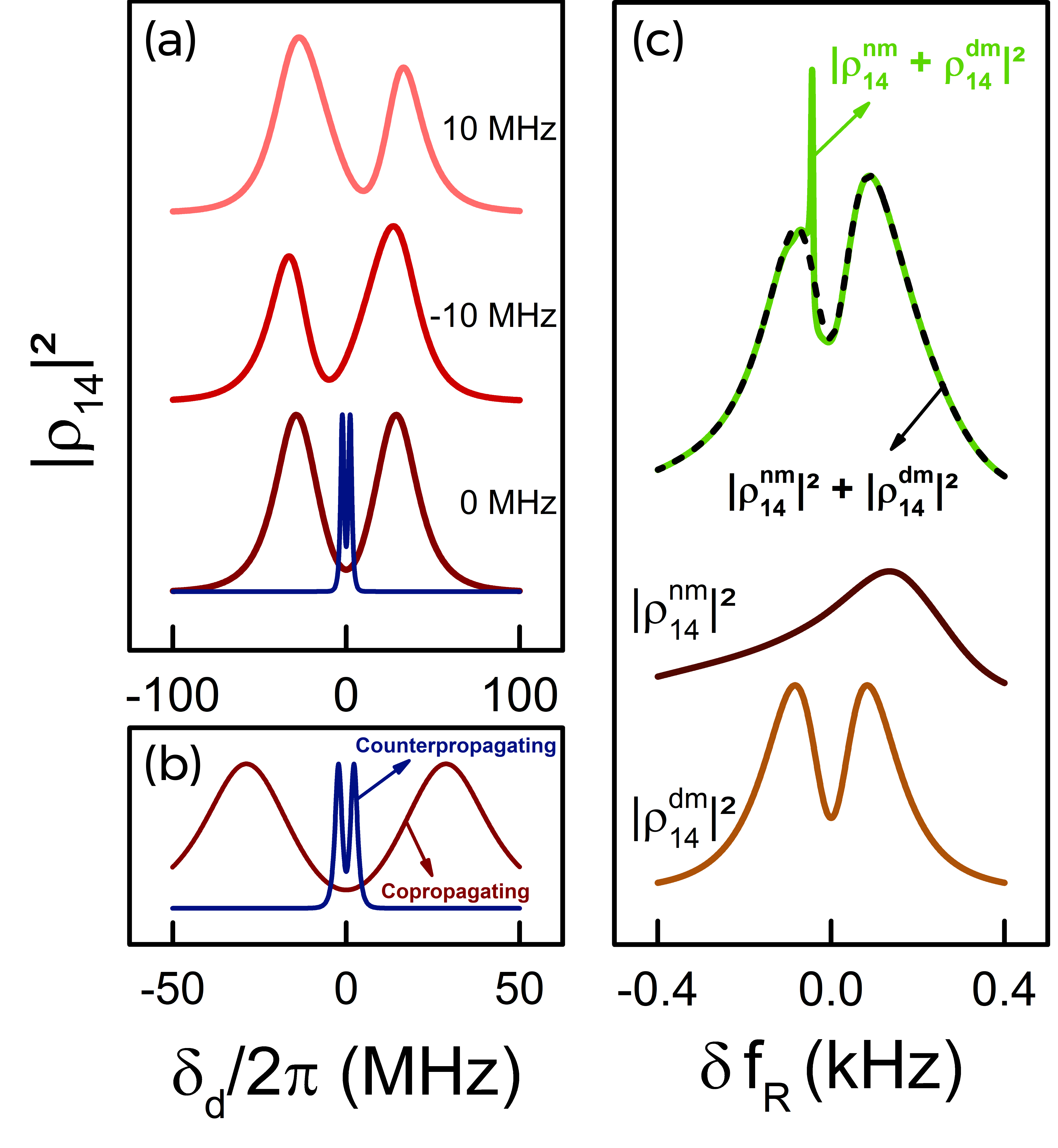}
\caption{(Color online) (a) $\left |\rho _{14} \right |^{2}$ scanning the diode laser frequency at a fixed repetition rate, for three different values of detuning of the $\omega_{m}$ mode regarding the transition $\left|2\right\rangle \rightarrow \left|3\right\rangle$. (b) Zoom of the bottom curve in (a) showing $\left |\rho _{14} \right |^{2}$ for co- and counter-propagating fields configuration. (c) Theoretical results scanning the repetition rate for a fixed diode frequency. $\left |\rho _{14} ^{dm} \right |^{2}$ is the result for the \textit{cw-fs} pathway, while $\left |\rho _{14} ^{nm} \right |^{2}$ accounts for the \textit{fs-fs} pathway considering the same repetition rate of Fig. 2 (b). The green and dashed curves are the two possible ways of adding these coherence pathways.}
\label{fig5}
\end{figure}

Our treatment of the problem begins with Liouville's equation with a electric dipole Hamiltonian as the interaction, 

\begin{equation}
\hat{H}_{int} = -\hslash\sum^4_{j \neq k}\left(\Omega_{l}e^{i\omega_{l}t}+c.c.\right)\left|j\right\rangle\left\langle k\right|,
\end{equation}

\noindent
where $\Omega_{l}$ and $\omega_{l}$ ($l = d, n, m$ or $i$) are the Rabi frequency and the optical frequency associated to the fields indicated in Fig. 4. Given the high repetition rate of our frequency comb, it can be treated as several cw lasers with a well defined frequency given by $\omega_{m}=2\pi(f_{0}+mf_{R})$ [$\omega_{n}=2\pi(f_{0}+nf_{R})$], where $f_{0}$ is the offset frequency and $m$ [$n$] is an integer number of order of 4 $\times$ $10^{5}$ . This grants the possibility of writing the Bloch equations as

\begin{equation}
\frac{\partial\rho_{jk}(t)}{\partial t} = -(i\omega_{jk} + \gamma_{jk})\rho_{jk}(t) - \frac{i}{\hslash}\left\langle j \right| [ \hat{H}_{int},\hat{\rho} ] \left| k \right\rangle,\\
\end{equation}

\noindent
for only one of the excitation routes and then adapt the scanning parameters to account for the other pathway. In these equations, $\gamma_{jk}$ is the decay rate of the density matrix element $\rho_{jk}$ and $\omega_{jk}$ is the frequency of the $\left|j\right\rangle$ $\rightarrow$ $\left|k\right\rangle$ transition. We also apply to our Bloch equations the rotating wave approximation and look for a steady-state solution. Propagation effects are neglected.

We solve the 16 coupled equations nonpertubatively by applying the cofactor expansion method on the system's matrix. Since we are looking for a generated signal between levels $\left|4\right\rangle$ and $\left|1\right\rangle$ of the theoretical model (see Fig. 4), our variable of interest is the coherence $\rho_{14}$. Once we have the full solution for this coherence, we integrate it over the Maxwell-Boltzmann distribution to account for the different velocity groups.

For the first experimental configuration, i.e. scanning the diode laser frequency we show the theoretical results in Fig. 5 (a). Each curve is presented for different detunings of the frequency comb mode with respect to the $\left|2\right\rangle$ $\rightarrow$ $\left|3\right\rangle$ transition. This means that for the two upper curves the laser exciting the second transition is not perfectly resonant with the atomic velocity group $v$ = 0. As a consequence, an asymmetry arises due to the Maxwell-Boltzmann distribution. We also show, in the bottom curve, the theoretical result when the diode beam propagates in an opposite direction to the fs beam (blue curve). A zoom of the theoretical results for these two fields configuration (co- and counter-propagating) is displayed in Fig. 5 (b). All the other parameters are the same. We can clearly see in this figure that the contribution of the Doppler effect for the AT splitting and the linewidth of the peaks are much larger for the copropagating configuration, in agreement with the present experiment performed by scanning the strong field.

This model was also used to reproduce the experimental curves presented in Fig. 2, changing only the Rabi frequency of the diode laser. Then, from each theoretical spectrum the AT splitting and the peak intensity were extracted. For this case, $f_0$ and $f_R$ were chosen in such a way that the mode of the frequency comb was assumed to be resonant with the $\left|2\right\rangle$ $\rightarrow$ $\left|3\right\rangle$ transition for atomic velocity group $v$ = 0. To avoid considering propagation effects, the experimental data is presented in terms of laser power measured at the entrance of the vapor cell. The theoretical data are plotted in Figs. 3(b) and (d) as a function of $|\Omega_{d}/\gamma_{22}|^2$ and scaled to compare with the experimental curves.

The outcomes of the configuration scanning the repetition rate for a fixed diode laser frequency are displayed in Fig. 5 (c). The two bottommost curves are the different excitation pathways, described previously. To obtain these curves we use exactly the same equations that produce the results in Fig. 5 (a), for a configuration of copropagating beams, but adapting the way the frequency is swept. To consider only the \textit{cw-fs} pathway, the field in the $\left|1\right\rangle$ $\rightarrow$ $\left|2\right\rangle$ transition is fixed. On the other hand, for the \textit{fs-fs} pathway, the first two fields are scanning, keeping the frequency difference given by the experimental repetition rate.

The \textit{cw-fs} pathway gives essentially the same result of Fig. 5 (a), when the repetition rate is fixed. The only palpable difference is in the AT splitting, which is slightly smaller but with same behavior of Fig. 3 (c). In contrast, the \textit{fs-fs} result is quite different. Since both fields that induce the first two transitions are scanning, a much broader spectrum arises. This is due to the AT effect caused by the frequency comb mode $\omega_{n}$, but blurred by the other scanning mode. The maximum intensity of the CBL generated in this case occurs when the system meets the double-resonance condition~\cite{Ban2013}:

\begin{equation}
	\label{double resonance}
	f_R = \dfrac{\omega_{23}-\omega_{12}}{2\pi(m-n)},
\end{equation}

\noindent
where \textit{m} and \textit{n} are integer numbers that determine a pair of comb modes that satisfies resonant condition for both excitation steps: $\left|1\right\rangle$ $\rightarrow$ $\left|2\right\rangle$ and $\left|2\right\rangle$ $\rightarrow$ $\left|3\right\rangle$.

In order to get the final signal, we must add the two possible coherence pathways. This operation can be done in two ways. Either we assume that these processes are independent, taking the square modulus of each coherence and adding them (dashed curve in Fig. 5 (c)) or we say that these coherences may interfere. If that is the case,  we must first add the coherences and then take the square modulus of this sum. This leads to the green curve on Fig. 5 (c), with the narrow peak over the AT doublet, just as it is in the experiment.

The narrow interference peak appears shifted from the resonance regarding the $v=0$ group. It does so because the frequency comb $\omega_{n}$ and the diode laser field must be simultaneously in resonance with the $\left|1\right\rangle$ $\rightarrow$ $\left|2\right\rangle$ transition for the exact same group of atoms. For the repetition rate used both in experiment and theory, this condition is satisfied for a group of atoms with some velocity.

It is possible, then, to shift the narrow peak by changing the repetition rate. These changes must be in a small range to ensure that there is a group of atoms interacting simultaneously with both lasers (the cw laser and two modes of the frequency comb) \cite{Moreno2011b}. If the change in repetition rate is 463.998 kHz (Eq. \ref{double resonance}), then another pair of modes will be able to produce exactly the same interference pattern, i.e. similar in position and amplitude. 

Likewise, it is possible to change the detuning of diode laser, assumed to be zero in Fig. 5 (c), to shift the narrow peak. This has been tested and also confirms that the interference process only occurs for those atoms simultaneously interacting with both lasers.

We remark that the linewidth of the observed narrow interference peak, of about 10 MHz, is mainly limited by experimental conditions such as the repetition rate scanning. On the other hand, the theoretical model shows that the peak has a linewidth of order of 1 MHz, a result that appears to be limited by the lifetime of the 5D$_{5/2}$ state \cite{Sheng}.

Since this is an interference process, phase must play an important role \cite{Jeong}. For the theoretical results here, the lasers were assumed to be on phase whilst in the experiment there was no specific phase control. By changing the phase on the theory it is possible to control the interference process ranging from a constructive to a destructive signal. This was not observed experimentally, for there was no phase control between lasers. A new experiment is being developed on which the lasers will have their phases locked. By carefully changing the optical path of the diode laser, we expect to see distinct behaviors of the interference process.

It is interesting to ask whether this type of interference process can also be observed using only cw lasers. In this context, a numerical analysis indicates that we would need at least three cw lasers, and to carefully engineer the frequency scan to vary simultaneously the frequency of two lasers at a rate that simulates the role played by the mode-locked femtosecond laser.

\section{Conclusions}
We have observed an interference effect between two four-wave mixing excitation routes involving a combination of cw laser and a 1 GHz frequency comb. This effect appears in the coherent blue light generated in rubidium vapor only when both lasers are resonant with one- and two- photon transitions for the same atomic velocity group. Remarkably, the signature of the interference is a narrow peak over an Autler-Townes doublet when we scan the repetition rate of the frequency comb. In particular, this process allows us to select and probe a specific velocity-group of atoms in a medium inhomogeneously broadened.

We also analyze the Autler-Townes splitting in a configuration of copropagating fields, scanning either the cw laser or the repetition rate of the frequency comb. In both situations, the cw laser induces the splitting. Furthermore, the experimental and theoretical results indicate that, under copropagating beams configuration, the Doppler effect plays an important role both in the Autler-Townes splitting and in the linewidth of the peaks.

A theoretical model in the frequency domain was proposed in order to describe these physical phenomena. For the AT splitting our model achieves good agreement with the experimental data, while for the interference process it provides evidence of the underlying mechanism giving rise to the effect. In fact, we were able to distinguish the role of each laser in both physical processes. Most importantly, we revealed that the combination of a cw laser and a 1 GHz frequency comb, with their respective characteristics, opens the possibility to induce different pathways that may lead to non-trivial quantum interferences. 

\vspace{5mm}

This work was supported by CAPES (PROEX 534/2018, No. 23038.003382/2018-39) and FAPERO (No. 01.1331.00031-00.057/2017). A. A. C. de Almeida acknowledges financial support by CNPq (132833/2017-4).

\end{document}